\documentclass[journal]{IEEEtran}
\usepackage{amsmath}
\usepackage{rotating}
\usepackage{amssymb}
\usepackage{graphicx}
\usepackage{makecell}
\usepackage{caption}
\usepackage{subfig}
\usepackage{relsize}
\usepackage[usenames, dvipsnames]{color}
\usepackage{booktabs}

\usepackage{enumitem}
\usepackage{booktabs}
\usepackage{mathtools}
\usepackage[english]{babel}
\usepackage{multirow}
\usepackage{lineno}
\usepackage[colorlinks = true,
            linkcolor = blue,
            urlcolor  = blue,
            citecolor = red,
            anchorcolor = blue]{hyperref}
            
\ifCLASSINFOpdf
\else
\fi
\begin{document}
%
\title{SleepEEGNet: Automated Sleep Stage Scoring with Sequence to Sequence Deep Learning Approach
}
%
%
%



\author{\IEEEauthorblockN{Sajad Mousavi, Fatemeh Afghah}\\
\IEEEauthorblockA{\textit{School of Informatics, Computing and Cyber Systems}, 
{Northern Arizona  University}, Flagstaff, USA \\ \{SajadMousavi,Fatemeh.Afghah\}@nau.edu}\\ 
\vspace{+0.2cm}
\IEEEauthorblockN{U. Rajendra Acharya}\\
\IEEEauthorblockA{\textit{Department of Electronics and Computer Engineering, Ngee Ann Polytechnic, Singapore} \\
\textit{Department of Biomedical Engineering, School of Science and Technology, Singapore University of Social Sciences, Singapore}\\
\textit{Department of Biomedical Engineering, Faculty of Engineering, University of Malaya, Malaysia} \\
aru@np.edu.sg} 
}

\maketitle

\begin{abstract}
Electroencephalogram (EEG) is a common base signal used to monitor brain activity and diagnose sleep disorders. Manual sleep stage scoring is a time-consuming task for sleep experts and is limited by inter-rater reliability. In this paper, we propose an automatic sleep stage annotation method called \textit{SleepEEGNet} using a single-channel EEG signal. The \textit{SleepEEGNet} is composed of deep convolutional neural networks (CNNs) to extract time-invariant features, frequency information, and a sequence to sequence model to capture the complex and long short-term context dependencies between sleep epochs and scores. In addition, to reduce the effect of the class imbalance problem presented in the available sleep datasets, we applied novel loss functions to have an equal misclassified error for each sleep stage while training the network. We evaluated the proposed method on different single-EEG channels (i.e., Fpz-Cz and Pz-Oz EEG channels) from the Physionet Sleep-EDF datasets published in 2013 and 2018. The evaluation results demonstrate that the proposed method achieved the best annotation performance compared to current literature, with  an overall accuracy of 84.26\%, a macro F1-score of 79.66\% and $\kappa$ = 0.79. Our developed model is ready to test with more sleep EEG signals and aid the sleep specialists to arrive at an accurate diagnosis. The source code is available at \url{https://github.com/SajadMo/SleepEEGNet}\footnote{This material is based upon work supported by the National Science
Foundation under Grant Number 1657260. Research reported in this publication was also supported by the National Institute On Minority Health And Health Disparities of the National Institutes of Health under Award Number
U54MD012388.}.

\end{abstract}

\begin{IEEEkeywords}
Sleep stage scoring, EEG analysis,
deep learning, sequence to sequence model.
\end{IEEEkeywords}

%
\IEEEpeerreviewmaketitle

\section{Introduction}
\label{sec:intro}
%
%
%
%
\IEEEPARstart{T}{he} electroencephalogram (EEG), electrooculogram (EOG), and electromyogram (EMG) signals are widely used to diagnose the sleep disorders (e.g., sleep apnea, parasomnias, and hypersomnia). These signals are typically recorded by placing sensors on different parts of the patient's body. In an overnight polysomnography (PSG) (also called as sleep study), usually EEG is mainly used to monitor the brain activities to diagnose sleep disorders \cite{biswal2017sleepnet} and other common disorders such as epilepsy \cite{acharya2018deep}.

The EEG signals are split into a number of predefined fixed length segments which are termed as epochs. Then, a sleep expert manually labels each epoch according to sleep scoring standards provided by the American Academy of Sleep Medicine (AASM) \cite{berry2012aasm} or the Rechtschaffen and Kales standard \cite{rechtschaffen1968manual}. Each EEG recording is around 8-hour long on average. Therefore, the manual scoring of such a long signal for a sleep expert is a tedious and time-consuming task. The human-based annotation methods also highly rely on an inter-rater agreement in place. Therefore, such restrictions call for automated sleep stage classification system that is able to score each epoch automatically  with high accuracy.

Several studies have focused on developing automated sleep stage scoring algorithms. Generally, they can be divided into two different categories in terms of the feature extraction approaches. First, the hand-engineered feature-based methods that require a prior knowledge of EEG analysis to extract the most relevant features. These approaches first extract the most common features such as time, frequency and time-frequency domain features \cite{miran2018real,zaeri2018feature,chen2017remote,chen2018predictive} of single channel-EEG waveforms. Then, they apply conventional machine learning algorithms such as support vector machines (SVM) \cite{koley2012ensemble}, random forests \cite{fraiwan2012automated,moghaddam2012learning} and neural networks \cite{hsu2013automatic} to train the model for sleep stage classification based on the extracted features. Although these methods have achieved a reasonable performance, they carry several limitations including the need for a prior knowledge of sleep analysis and not able to generalize to larger datasets from various patients with different sleep patterns.
Second, automated feature extraction-based methods such as deep learning algorithms in which the machine extracts the pertaining features automatically (e.g. CNNs to extract time-invariant features from raw EEG signals).

In recent years, deep neural networks have shown impressive results in various domains ranging from computer vision and reinforcement learning to natural language processing \cite{sutskever2014sequence,mousavi2016deep,mousavi2016learning,mousavi2017traffic,mousavi2017applying,mousavi2014automatic,shamsoshoara2019distributed}. One key reason for the success of deep learning based  methods in these domains is the availability of large amounts of data to learn the underlying complex pattern in the data sets. Due to availability of large sleep EEG recordings \cite{goldberger2000physiobank}, deep learning algorithms have also been applied for sleep stage classification \cite{biswal2017sleepnet,yildirim2019deep,supratak2017deepsleepnet,tsinalis2016automatic,michielli2019cascaded} and other. However, in spite of the remarkable achievements in using deep learning models compared to the shallow machine learning methods for sleep stage scoring task, they still suffer from the class imbalance problem present in the sleep datasets. Thus, this problem limits the use of deep learning techniques and in general machine learning techniques for the sleep stage classification and to reach an expert-level performance for the sleep stage classification. 

In this study, we introduced a novel deep learning approach, called \textit{SleepEEGNet}, for automated sleep stage scoring using a single-channel EEG. We believe the sleep stage classification problem is sequential in nature. Therefore, we applied a sequence to sequence deep learning model with the following building blocks: (1) CNNs to perform the feature extraction, (2) bidirectional recurrent neural network (BiRNN) to capture temporal information from sequences and consider the previous and future input information simultaneously, and (3) attention network to let the model learn the most relevant parts of the input sequence while training. Also, we utilized new loss functions to reduce the effect of imbalance class problem on the model by treating the error of each misclassified sample equally regardless of being a member of the majority class or minority class.


The rest of the paper is structured as follows: Section \ref{sec:propsed} describes the proposed method. Section \ref{sec:expres} presents data set and data preparation, the experimental design, and shows the achieved results by the proposed method along with a performance comparison to the state-of-the-art algorithms. Finally, Section \ref{sec:con} concludes the paper.


\section{Methodology}
\label{sec:propsed}
In the following sections, we present a detailed description of our proposed novel model developed to automatically score each sleep stage from a given EEG signal. 

\subsection{Pre-processing}
The input to this method is a sequence of 30-s EEG epochs. In order to extract the EEG epochs from a given EEG signal, we follow two simple steps:

\begin{enumerate}[noitemsep,nolistsep]
    \item Segmenting the continuous raw single-channel
    EEG to a sequence of 30-s epochs and assigning a label to each epoch (i.e., sleep stage) based on the annotation file. 
	\item Normalizing 30-s EEG epochs such that each one has a zero mean and unit variance.
\end{enumerate}
 It is worth mentioning that, these pre-processing steps for the sleep stage extraction are very simple and do not involve any form of filtering or noise removal methods.

\begin{figure*}[htb]
\centering
  \includegraphics[height=0.4\textheight,width=\linewidth]{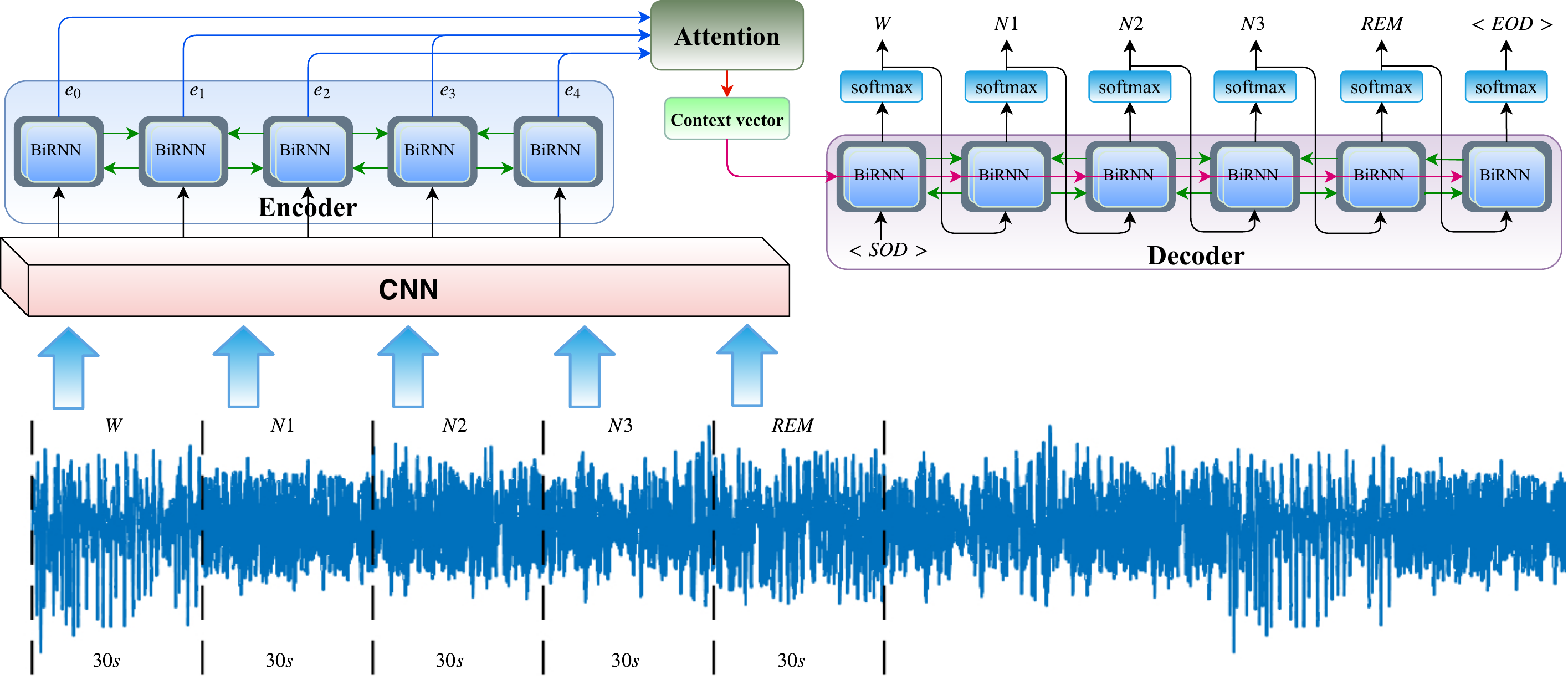}
  \caption{Illustration of sequence to sequence deep learning network architecture used for automated sleep stage scoring.} 
  \label{fig:final-model}
\end{figure*}

\subsection{The architecture}
The sequence to sequence models have shown very impressive results in neural machine translation applications, nearly similar to human-level performance \cite{johnson2016google}. 
The architecture of sequence to sequence networks is usually composed of two main parts: the encoder and decoder which are types of recurrent neural network (RNN). In this study, we used an RNN sequence to sequence model along with a convolutional neural network (CNN) to perform automatic sleep stage
scoring task.

Figure \ref{fig:final-model} illustrates the proposed network architecture  for automatic sleep stage classification.
We applied almost the same CNN architecture provided by \cite{supratak2017deepsleepnet}. The CNN  consists of two sections, one with small filters to extract temporal information and another with large filters to extract frequency information. The idea behind these variable sizes of filters comes from the signal processing community to have a trade-off between extracting time domain and frequency domain features \cite{cohen2014analyzing}. This helps get benefit from both time and frequency domain features in the classification task.  Each CNN part consists of four consecutive one-dimensional convolutional layers. Each convolutional layer is passed to a rectified linear unit (ReLU) nonlinearity. The first layer is followed by a max pooling layer and a dropout block, and just a dropout block comes after the last convolutional layer. At each time-step of training/testing the model, a sequence (size of $maxtime$) of 30-s EEG epochs is fed into the CNN for feature extraction. In the end, the outputs of CNN parts are concatenated and followed by a dropout block in order for the encoder network to encode the sequence input.
 Figure \ref{fig:cnn_part} depicts the detailed CNN structure.


The sequence to sequence model is designed based on the encoder-decoder abstract ideas. The encoder encodes the input sequence, while the decoder computes the category of each single channel 30-s EEG of the input sequence. The encoder is composed of long short-term memory (LSTM) units which captures the complex and long short-term context dependencies between the inputs and the targets \cite{fernandez2015using}. They capture non-linear dependencies present in the entire time series while predicting a target. The (time) sequence of input feature vectors herein are fed to the LSTMs and then the hidden states, $(e_0, e_1, e_2, \ldots)$, calculated by the LSTM are considered as the encoder representation, and are fed  to the attention network (or to initialize the first hidden state of the decoder, if the basic decoder is used), as depicted in Figure \ref{fig:final-model}.

\subsection{Bidirectional recurrent neural network}
\begin{figure}[htb]
\centering
  \includegraphics[height=0.4\textheight,width=0.6\linewidth,keepaspectratio]{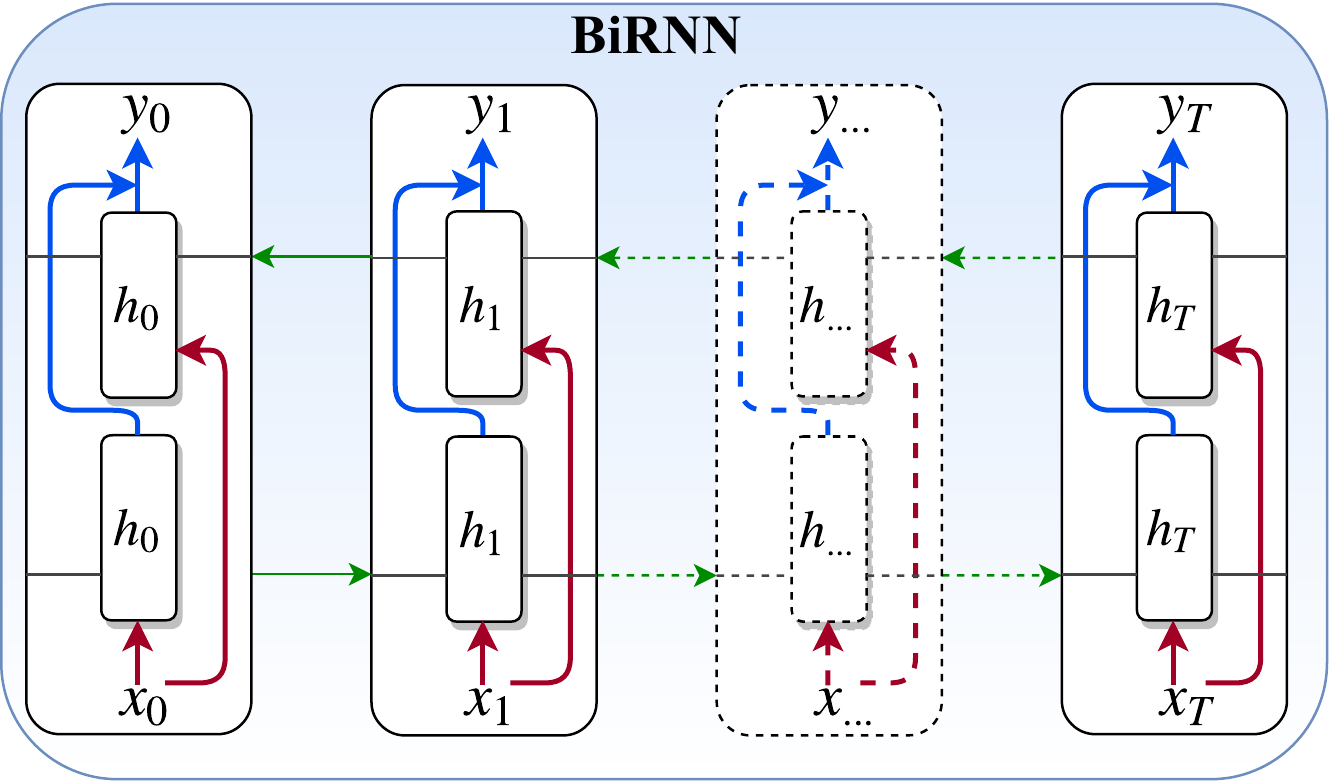}
  \caption{A schematic diagram of the bidirectional recurrent neural network.} 
  \label{fig:rnn_part}
\end{figure}

We have utilized the bidirectional recurrent neural network (BiRNN) units in the network architecture instead of the standard LSTM (i.e., standard RNN). Standard RNNs are unidirectional, hence they are restricted to use the previous input state. To address this limitation, the BiRNN have been proposed \cite{schuster1997bidirectional}, which can process data in both forward and backward directions. Thus, the current state has access to previous and future input information simultaneously. The BiRNN consists of forward and backward networks. The input sequence is fed in normal time order, $t = 1, ...,T$ for the forward network, and in reverse time order, $t = T, ..., 1$ for the backward network. Finally, the weighted sum of the outputs of the two networks is computed as the output of the BiRNN. This mechanism can be formulated as follow:

\begin{figure}[ht]
\centering
  \includegraphics[height=0.7\textheight,width=0.9\linewidth,keepaspectratio]{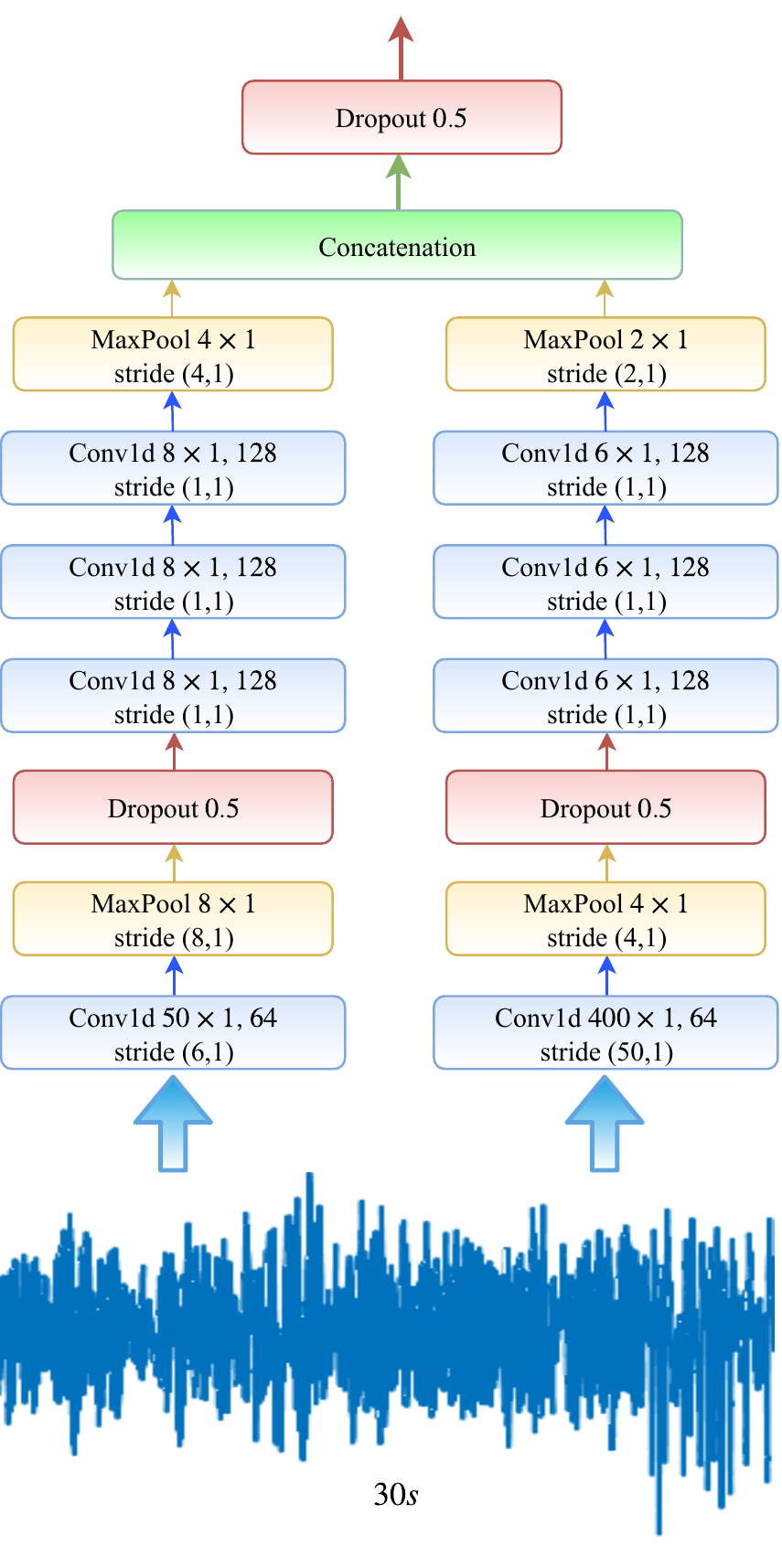}
  \caption{Detailed sketch of the used CNN model used in the proposed work.} 
  \label{fig:cnn_part}
\end{figure}

\begin{align}
  &\begin{aligned} 
    \overrightarrow{h_t}= \tanh(\overrightarrow{W}x_t+\overrightarrow{V}{\overrightarrow{h}}_{t-1}+\overrightarrow{b})
  \end{aligned}\\
  &\begin{aligned}
   \overleftarrow{h_t}= \tanh(\overleftarrow{W}x_t+\overleftarrow{V}{\overleftarrow{h}}_{t+1}+\overleftarrow{b})
  \end{aligned} \\
    &\begin{aligned}
  y_t = (U[ \overrightarrow{h_t}; \overleftarrow{h_t}] + b_y),
  \end{aligned}
\end{align}
where ($\overrightarrow{h_t}$, $\overrightarrow{b}$) are the hidden state and the bias of the froward network, and ($\overleftarrow{h_t}$, $\overleftarrow{b}$) are the hidden state and the bias of the backward network. Also, $x_t$ and $y_t$ are the input and the output of the BiRNN, respectively. Figure \ref{fig:rnn_part} illustrates a BiRNN architecture with T time steps.

\subsection{Attention Decoder}
The decoder is used to generate the target sequence epoch by epoch. Similar to the encoder, the building block of the decoder is an LSTM. In a basic decoder, at every step of decoding, the decoder gets a new representation of an input element of the sequence generated by the encoder and an element of the target input. The last element of the input sequence is usually the last influence to update the encoder's hidden state. Therefore, the model can be biased to the last element. To address such a problem, we have applied an attention mechanism to the model to consider not only the whole encoder representation of the input but also it can learn to put more emphasis on different parts of the encoder outputs in each step of decoding. In other words, the attention mechanism makes the model to learn the most relevant parts of the input sequence in the decoding phase. In a sequence to sequence model without attention approach, the decoder part relies on the hidden vector of the decoder's RNN (or BiRNN), while the sequence to sequence model with the attention is more purposeful. It considers the combination of encoder's representation and decoder hidden vector calling the context vector or the attention vector, $(c_t)$. 

To calculate the $c_t$ vector, we first computed a set of attention weights with a function $f(.)$. These are probabilities, $(\alpha_i)$, corresponding to the importance of each hidden state. Then, these scores are multiplied by the hidden states (i.e, the encoder output vectors) to calculate the weighted combination, $(c_t)$. 

\begin{align}
  &\begin{aligned}
   f(h_{t-1},e_i) = \tanh({W_h} h_{t-1}+{W_e} e_i)
  \end{aligned}\\
  &\begin{aligned}
    \alpha_{i} & = softmax(f(h_{t-1},e_i)) \approx  \frac{\exp(f(h_{t-1},e_i))}{\sum_{j=1}^{n} \exp(f(h_{t-1},e_j))} \\
      & i \in 1,2,\ldots ,{n},
  \end{aligned} \\
    &\begin{aligned}
   c_t & =\sum_{i=0}^{n} \alpha_{i} e_{i},
  \end{aligned}
\end{align}
where $\alpha_{i}$ is the signification of the part $i$ of hidden state. In other words, at every time step $t$, the attention layer computes $f(.)$, a combination of the values of $e_{i}$ (the encoder's hidden state) and $h_{t-1}$ (the decoder's hidden state) followed by a $\tanh$ layer. Then, it is fed into a softmax module to calculate $\alpha_{i}$ over ${n}$ parts. Finally, the attention module computes $c_t$, a weighted sum of all  vectors $e_{i}$ based on computed $\alpha_{i}$'s. Thus, the model can learn to focus on the important regions of the input sequence when decoding.

During the training phase, the decoder, in addition to the augmented version of the encoder's hidden states, captures the given target sequence shifted by one starting with a special feature vector $<SOD>$ (i.e., the start of decoding) as input. Then, the decoder starts to generate outputs until it confronts the special label called $<EOD>$ (i.e., the end of decoding). We should note that the target sequence is just used during the training phase and is not applied for the testing phase. During the testing phase, the decoder uses whatever label it generates at each step as the input for the next step. Finally, a softmax is applied to the output of the decoder to convert it to a vector of probabilities $p \in \mathbb{R}^C$, where $C$ represents the number of classes and each element of $p$ indicates the probability of each class of the sleep stage.

\subsection{Loss calculation}
\label{sec:loss}
Similar to other biomedical applications, the sleep stage classification also deals with the problem of class imbalance. To alleviate the effect of this problem on the model, we calculated new loss functions based on \cite{wang2016training} to treat the error of each misclassified sample equally regardless of being a member of the majority or minority class.

We extended the proposed loss functions, mean false error (MFE) and mean squared false error (MSFE), in \cite{wang2016training} for the multi-class classification task. MFE and MSFE can be defined as follows:
\begin{align}
 &\begin{aligned} 
l(c_i)= \frac{1}{C_i} \sum_{i=j}^{C_i} (y_j-\hat{y_j})^2,
  \end{aligned}\\
 &\begin{aligned} 
l_{MFE}= \sum_{i=1}^{N} l(c_i),
  \end{aligned}\\
 &\begin{aligned} 
l_{MSFE}= \sum_{i=1}^{N} l(c_i)^2,
  \end{aligned}
\end{align}
where $c_i$ is the class label (e.g., W or N1), $C_i$ is the number of the samples in class $c_i$, $N$ is the number available classes (here sleep stage classes). and $l(c_i)$ is the calculated error for the class class $c_i$. In the most common used loss function, mean squared error (MSE), the loss is calculated by averaging the squared difference between predictions and targets. This way of computing the loss makes the contribution of the majority classes be much more in comparison with the minorities classes in the imbalanced dataset. However, the MFE and MSFE try to consider the errors of all classes equally. 

\section{Experimental Results}
\label{sec:expres}
\subsection{Dataset and Data Preparation}
In this study, we used 
the Physionet Sleep-EDF dataset \cite{kemp2000analysis,goldberger2000physiobank}: version 1 contributed in 2013 with 61 polysomnograms (PSGs) and version 2 contributed in 2018 with 197 PSGs to evaluate the performance of the proposed method for the sleep stage scoring task. The Sleep-EDF dataset contains two different studies including (1) study of age effects on sleep in healthy individuals (SC = Sleep Cassette) and (2) study of temazepam effects on sleep (ST = Sleep Telemetry).
The dataset includes whole-night polysomnograms (PSGs) sleep recordings at the sampling rate of 100 Hz. Each record contains EEG (from Fpz-Cz and Pz-Oz electrode locations), EOG, chin electromyography (EMG), and event markers. Few records often also contain oro-nasal respiration and rectal body temperature. The hypnograms (sleep stages; 30-s epochs) were manually labeled by well-trained technicians according to the Rechtschaffen and Kales standard \cite{rechtschaffen1968manual}. Each stage was considered to belong to a different class (stage). The classes include W, REM, N1, N2, N3, N4, M (movement time) and '?' (not scored). According to American Academy of Sleep Medicine (AASM) standard, we integrated the stages of N3 and N4 in one class named N3 and excluded M (movement time) and ? (not scored) stages to have five sleep stages \cite{berry2012aasm}. In addition, we considered Fpz-Cz/Pz-Oz EEG channels from SCs of both versions in our evaluations. Table \ref{tab:StatDataset} presents the   number of sleep stages in two different versions.

\begin{table}[!ht]
\centering{

    \caption{Details of number of sleep stages in each version of Sleep-EDF dataset.}
    \label{tab:StatDataset} 
    \resizebox{0.8\linewidth}{!}{ 
    \begin{tabular}{l|c|c|c|c|c|c} 
    \toprule[\heavyrulewidth]
\hline
  
      \textbf{Dataset} & \textbf{W} & \textbf{N1} &  \textbf{N2} &\textbf{N3-N4} & \textbf{REM} & \textbf{Total}\\ 
       \hline
      Sleep-EDF-13 & 8,285 & 2,804 & 17,799&5,703&7,717&42,308\\
           Sleep-EDF-18 & 65,951 & 21,522 & 96,132&13,039&25,835&222,479\\

     \hline
      \bottomrule
    \end{tabular}
    }
}
\end{table}

\subsection{Experimental Design}
The distribution of sleep stages in the Sleep-EDF database is not uniform. Hence, the number of W and N2 stages are much greater than other stages. The machine learning approaches do not perform well with the class imbalance problem. To address this problem, in addition to using the novel loss functions described in Section \ref{sec:loss}, the dataset is oversampled to nearly reaching a balanced number of sleep stages in each class. We have used the synthetic minority over-sampling technique (SMOTE) to generates the synthetic data points by considering the similarities between existing minority samples \cite{chawla2002smote}.

Our proposed model was evaluated using k-fold cross-validation. We set k to 20 and 10 for version 1 and version 2 of the Sleep-EDF dataset, respectively. In other words, we split the dataset into k folds. Then, for each unique fold, (1) fold is taken as test set and the remaining folds as a training set and (2) trained the model using the training set and evaluated the model using the test set. Finally, all evaluation results were combined.   

The network was trained (for each dataset) with a maximum of 400 epochs. RMSProp optimizer was used to minimize the $l_{MFE}$ loss with mini batches of size $20$ and a learning rate of $\alpha = 0.0001$. We also applied an additional $L_2$ regularization element with $\beta = 0.001$ to the loss function to mitigate the overfitting.  Python programming language  and Google Tensorflow deep learning library were utilized to implement our proposed approach.
\subsection{Evaluation Metrics}
We have used different metrics to evaluate the performance of the proposed approach including, overall accuracy, precision, recall (sensitivity), specificity, Cohen's Kappa coefficient ($\kappa$) and  F1-score. We also computed macro-averaging of F1-score (MF1) which is the sum of per-class F1-scores over the number of classes (i.e., sleep stages).


\begin{table} [ht]  

\caption{Confusion matrix and per-class performance achieved by the proposed method using Fpz-Cz EEG channel of the EDF-Sleep-2013 database.}
\renewcommand{\arraystretch}{1.4}
 \centering{
\label{tab:cmfpz-cz13}
	\resizebox{0.85\linewidth}{!}{  
\begin{tabular}{cccccccccccc}
 \toprule
\textbf{} & \textbf{} &  \multicolumn{5}{c} {Predicted} & \multicolumn{4}{c} {Per-class Performance (\%)} \\
 \cmidrule(lr){3-7} 
 \cmidrule(lr){8-11}
\textbf{} & \textbf{}& W1 & N1& N2&N3&REM & Pre &Rec&Spe&F1 \\
\cline{2-11}
\multirow{5}{*}{\begin{turn}{90}Actual\end{turn}}  &\multicolumn{1}{|l|}{W1}
& 7161  & 432  &  67  &  27 &  219 & 87.84 &90.58 &96.97 &89.19
 \\ 
&\multicolumn{1}{|l|}{N1}
& 442 & 1486  & 364  &  25  & 409 &50.05 &54.51 &96.08 &52.19
 \\ 
 &\multicolumn{1}{|l|}{N2}
& 359  & 735 &14187 & 1035 &  837 &91.26 &82.71 &94.20 &86.77
 \\ 
  &\multicolumn{1}{|l|}{N3}
&  37    & 9  & 560 & 4857  &   2 &81.69 &88.87 &96.90 &85.13
 \\ 
  &\multicolumn{1}{|l|}{REM}
& 153   & 307   & 368   &   2  & 6520 &81.63 &88.71 &95.59 &85.02
 \\ 
  \bottomrule  
\end{tabular} }
}

\end{table}

\begin{table} [ht]  

\caption{Confusion matrix and per-class performance achieved by the proposed method using Pz-Oz EEG channel of the EDF-Sleep-2013 database.}
\renewcommand{\arraystretch}{1.4}
 \centering{
\label{tab:cm-pz-oz13}
	\resizebox{0.85\linewidth}{!}{  
\begin{tabular}{cccccccccccc}
 \toprule
\textbf{} & \textbf{} &  \multicolumn{5}{c} {Predicted} & \multicolumn{4}{c} {Per-class Performance (\%)} \\
 \cmidrule(lr){3-7} 
 \cmidrule(lr){8-11}
\textbf{} & \textbf{}& W1 & N1& N2&N3&REM & Pre&Rec&Spe&F1 \\
\cline{2-11}
\multirow{5}{*}{\begin{turn}{90}Actual\end{turn}}  &\multicolumn{1}{|l|}{W1}
& 7094 &  398 &   82   & 41 &  238 & 90.20 &90.33 &97.65 &90.27
 \\ 
&\multicolumn{1}{|l|}{N1}
&  539 & 1167 &  455   & 29  & 492 &45.84 &43.51 &96.36 &44.64
 \\ 
 &\multicolumn{1}{|l|}{N2}
&  114 &  655& 14220 & 1157  & 971 &88.58 &83.07 & 92.19&85.74
 \\ 
  &\multicolumn{1}{|l|}{N3}
&   17 &   12 &  791 & 4658  &  10 &78.48 &84.88 &96.36 &81.55
 \\ 
  &\multicolumn{1}{|l|}{REM}
&  100  & 314  & 506  &  50  &6489 &79.13 &87.00 &94.84 &82.88
 \\ 
  \bottomrule  
\end{tabular} }
}
\end{table}

\begin{table*} [htb] 
\caption{Comparison of performance obtained by our approach with other state-of-the-art algorithms.}
\renewcommand{\arraystretch}{1.4}
 \centering{
\label{tab:comps}
	\resizebox{1.\linewidth}{!}{  
\begin{tabular}{ *{13}{c} }
\toprule[\heavyrulewidth]\toprule[\heavyrulewidth]
&Method    & Dataset & CV
            & EEG Channel
                    & \multicolumn{3}{c}{Overall Performance}
                            & \multicolumn{5}{c}{Per-class Performance (F1)}                \\    
                            \cmidrule(lr){6-8}
                             \cmidrule(lr){9-13}
                           &\textbf{} &\textbf{}&  \textbf{}&  \textbf{}& ACC & MF1& $\kappa$& W&N1&N2&N3&REM\\
                   
    \midrule
    
\multirow{8}{*}{\begin{turn}{90}Inter.\end{turn}} &\multicolumn{1}{|c}{\textit{SleepEEGNet}}   &   Sleep-EDF-13  & 20-fold CV &  Fpz-Cz  &   \textbf{84.26}  &   \textbf{79.66}  &   \textbf{0.79}  &   \textbf{89.19}  &   \textbf{52.19}  &   \textbf{86.77} &   \textbf{85.13}&   \textbf{85.02}  \\
   &\multicolumn{1}{|c}{ Supratak et al. \cite{supratak2017deepsleepnet}}    &  Sleep-EDF-13  & 20-fold CV &  Fpz-Cz    &   82.0  &   76.9  &  0.76  &  84.7  &   46.6  &   85.9 &   84.8&  82.4  \\
   
    &\multicolumn{1}{|c}{ Tsinalis et al. \cite{tsinalis2016automatic}}    &  Sleep-EDF-13  &  20-fold CV & Fpz-Cz    &   78.9 &   73.7  &  -  &  71.6  &   47.0  &   84.6 &   84.0&  81.4 \\
        &\multicolumn{1}{|c}{ Tsinalis et al. \cite{tsinalis2016automatic2}}    &  Sleep-EDF-13  & 20-fold CV &  Fpz-Cz    &   74.8  &   69.8  &  -  &  65.4  &  43.7  &   80.6 &   84.9&  74.5  \\
     \cline{2-13}
    &\multicolumn{1}{|c}{\textit{SleepEEGNet}}    &  Sleep-EDF-13  & 20-fold CV &  Pz-Oz    &  \textbf{82.83}  &   \textbf{77.02}  &   \textbf{0.77}  &   \textbf{90.27}  &   \textbf{44.64}  &  \textbf{ 85.74} &  \textbf{ 81.55}&   \textbf{82.88}  \\
  &\multicolumn{1}{|c}{ Supratak et al. \cite{supratak2017deepsleepnet}}    &  Sleep-EDF-13  & 20-fold CV &  Pz-Oz    &   79.8  &   73.1  &   0.72  &   88.1  &   37  &   82.7 &   77.3&   80.3  \\
   \cline{2-13}
 &\multicolumn{1}{|c}{\textit{SleepEEGNet}}   &   Sleep-EDF-18  &  10-fold CV & Fpz-Cz  &   \underline{80.03}  &   \underline{73.55}  &   \underline{0.73}  &   \underline{91.72}  &  \underline{44.05}  &   \underline{82.49} &   \underline{73.45}&   \underline{76.06}  \\
 &\multicolumn{1}{|c}{\textit{SleepEEGNet}}    &  Sleep-EDF-18  &  10-fold CV & Pz-Oz    &   \underline{77.56}  &   \underline{70.00}  &   \underline{68.94}  &   \underline{90.26}  &   \underline{42.21}  &   \underline{79.71} &   \underline{94.83}&   \underline{72.19}  \\
 
      
 \bottomrule 
 \multicolumn{12}{c}{Sleep-EDF-13: Sleep-EDF 2013; Sleep-EDF-18: Sleep-EDF 2018; CV: Cross Validation}
\end{tabular}}}
\end{table*}
\subsection{Results and Discussion}

Tables \ref{tab:cmfpz-cz13} and \ref{tab:cm-pz-oz13} present the confusion matrices and the performances of each class achieved by the proposed method using   Fpz-Cz and Pz-Oz  channels  of  the  EDF-Sleep-2013 data set, respectively. The main diagonals in each confusion matrix denote the true positive (TP) values  which indicate the number of stages scored correctly. It can be seen from the tables (the confusion matrices' parts) that TP values are  higher than other values in the same rows and columns. These tables also show the prediction performance (i.e., the precision, recall, specificity and F1 score) of our model for each class (i.e., the stage). Among all stages, the model performance is better for W1, N2, N3, and REM stages than the N1 stage. This may be because the number of N1 stages in the dataset is smaller compared to the other stages. However, our results for stage N1 is better than other state-of-the-art algorithms listed in Table \ref{tab:comps}.

\begin{figure}[h]%
    \centering
    \subfloat[]{{\includegraphics[width=8.5cm]{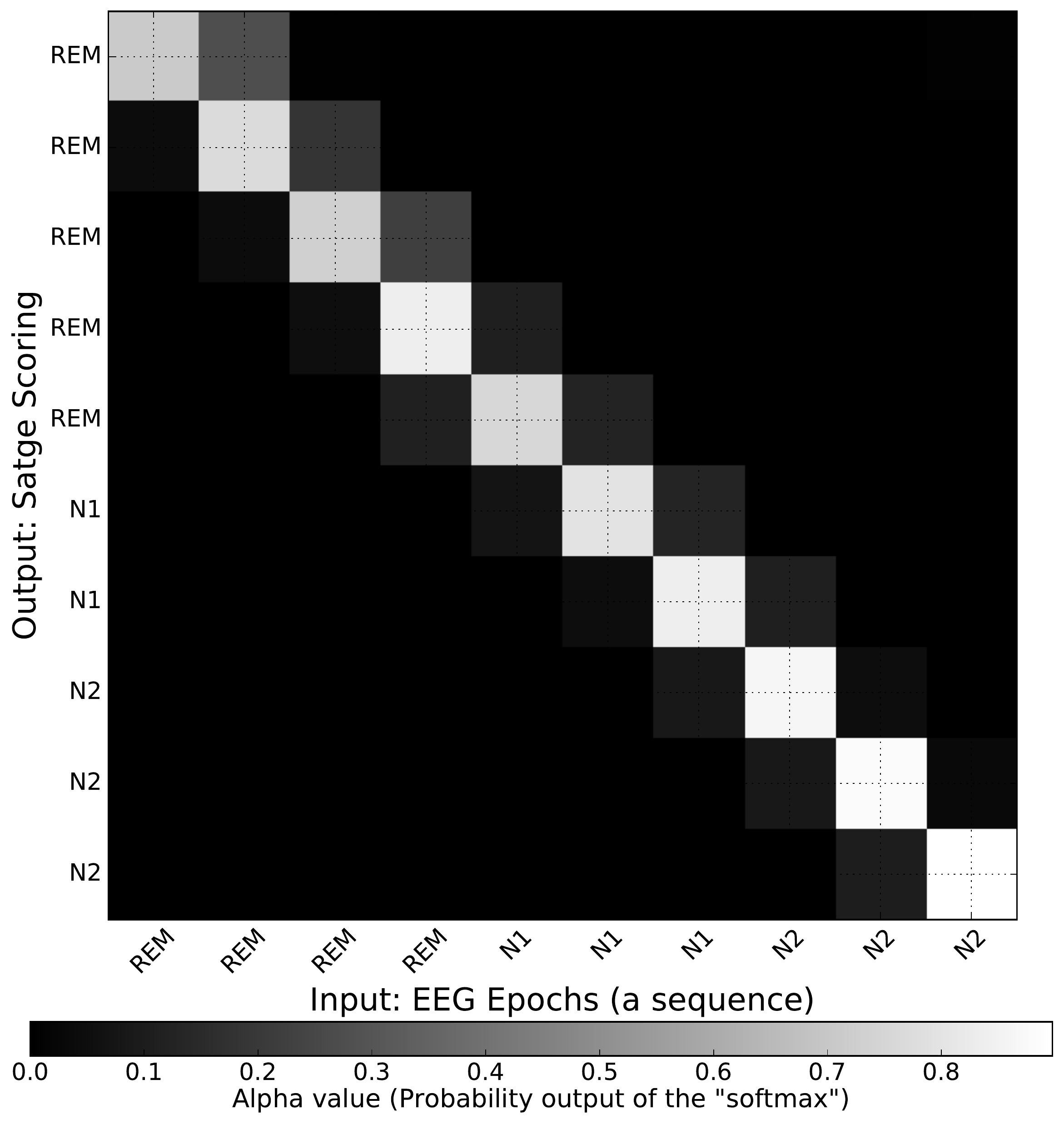} }}%
    \qquad
    \subfloat[]{{\includegraphics[width=8.5cm]{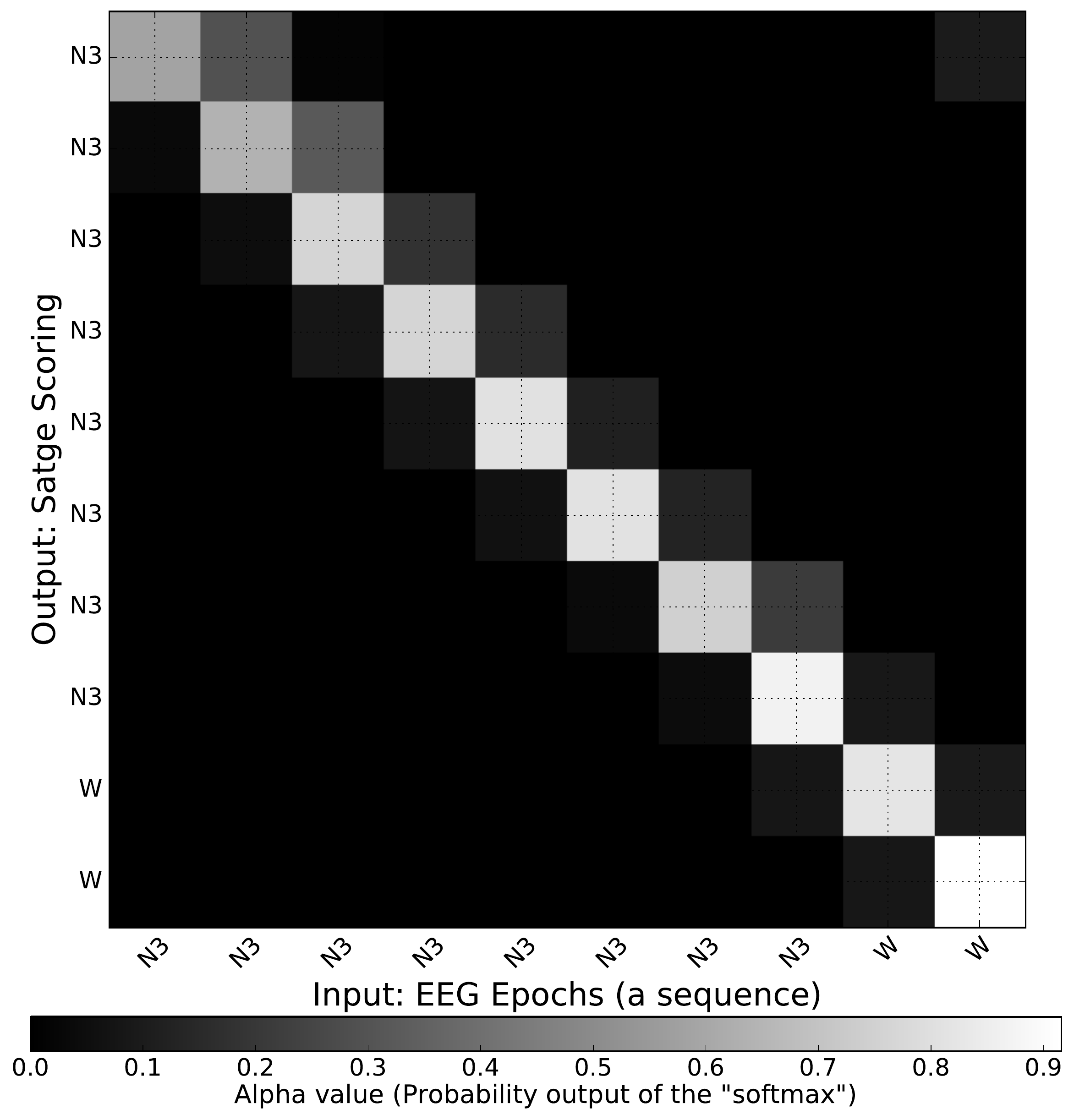} }}%
    \caption{Attention maps of two sequence inputs (EEG epochs) and their corresponding sleep stage scores provided by our proposed method.}%
    \label{fig:atten_map}%
\end{figure}


Typically, there are two approaches to evaluate the proposed methods in the literature: (i) intra-patient paradigm in which the training 
and evaluation sets can include epochs from the same subjects, and (ii) inter-patient paradigm in which the epochs sets for test and
training come from different individuals. As the inter-patient scheme seems to be a more realistic evaluation mechanism, the results and comparisons presented in this study are based on the inter-patient paradigm. Table \ref{tab:comps} presents the comparison of stage sleep scoring results for the proposed method with the existing algorithms. It can be noted from Table \ref{tab:comps} that the proposed model outperformed the state-of-the-art algorithms presented in the table. Our model has performed better in all listed channels (i.e., the Fpz-Cz and the Pz-Oz EEG channels) in terms of all evaluation metrics compared to others. It may be noted that in spite of the imbalance-class problem, our model yielded desirable performance, especially for stage N1. In addition to the Sleep-EDF 2013 dataset, we also evaluated our model with the Sleep-EDF 2018 dataset. Since the dataset has been published recently, we could not find any work to compare the performance of our model. Therefore, we just reported our findings without any comparison.  
 \begin{figure}[h]
\centering
  \includegraphics[width=1.\linewidth]{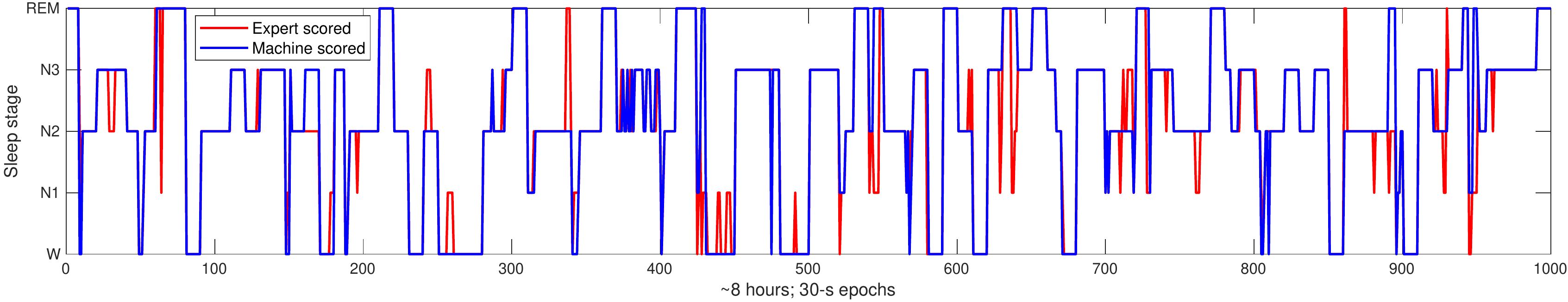}
  \caption{A example of hypnograms generated by the machine (i.e., the proposed method) and a sleep expert of a subject from the Sleep-EDF-13 dataset; approximately $85\%$ coverage.} 
  \label{fig:hyponogram}
\end{figure}

Figure \ref{fig:hyponogram} also illustrates the hypnogram produced manually by a sleep expert and its corresponding hypnogram generated by our method for a subject for approximately 8 hours of sleep at night. It can be noted from the figure that around 85\%  the manually scored hypnogram and automatically scored correctly.

Furthermore, by employing the attention mechanism into the network, we are able to illustrate (in the form of attention maps) which input epochs are important to score the sleep stages. As shown in Figure \ref{fig:atten_map}, we can see the network used almost the exact input epoch to predict its corresponding sleep stage.    

Our model has performed better than the rest of the works due to the following two reasons: First, the nature of the sleep stage scoring task is sequential in which each sleep stage has a relationship with the previous and next stage. Hence, applying a sequence to sequence deep learning model for such a problem would be a desirable choice. Also, using the attention model and BiRNNs as building blocks of the sequence to sequence model increased the performance. Second, the sleep stage classification suffers from the imbalance-class problem. To reduce the effect of this problem, we applied new loss functions (i.e., the MFE and MSFE) to have an equal misclassified error effect for each sleep stage while training the network.

One of the remarkable aspects of our proposed method is that, the model is generic in nature so it generalizes for other problems in the biomedical signal processing area that are inherently sequential and have the imbalance-class problem such as the heartbeat classification for arrhythmia detection \cite{mousavi2018inter,mousavi2018ecgnet}.

Even though our proposed model achieved significant results compared to the existing methods for the sleep stage classification, the model still carries several limitations. First, similar to other deep learning methods, our method needs a sufficient amount of sleep stage samples in training phase to learn discriminative features of each stage. Second, as our model is a sequence to sequence approach, at each time step, it requires to have a certain amount of 30-s EEG epochs (as input sequence) to be able to score the input epochs. Finally, our proposed method is evaluated with two available EEG channels (i.e., Fpz-Cz and Pz-Oz EEG channels) extracted from the Physionet Sleep-EDF datasets. Therefore, to evaluate its performance on other EEG channels, the network has to be trained with new EEG epochs.

Furthermore, in future, we intend to extend this work using multimodal polysomnography (PSG) signals including EEG, EOG (electrooculography) and EMG (electromyogram) to boost the performance of the sleep stage classification. 

\section{Conclusion}
\label{sec:con}

We have presented a novel and state-of-the-art algorithm for automated sleep stage annotation problem. The proposed method leverages the ability of deep convolutional neural network and encoder-decoder network in which we have used bidirectional recurrent neural networks and attention mechanisms as its building blocks. The proposed new loss calculation approaches helped to reduce the effect of the class-imbalance problem and boost the performance, especially the performance of our method on the stage N1, that is more difficult than other sleep stages to score. Table \ref{tab:comps} presents that, our proposed model significantly outperformed the existing algorithms by yielding the highest performance for the sleep stage scoring task. While developing the automated systems, generally there will be imbalance data problem (normal class more data than diseased class). Our developed model can be applied to such biomedical applications like arrhythmia detection using ECG signals, epilepsy detection using EEG signals and EMG signals to study the postures.


%




\ifCLASSOPTIONcaptionsoff
  \newpage
\fi

\bibliographystyle{IEEEtran}
\bibliography{bare_jrnl}
\end{document}